\documentclass[useAMS,usenatbib]{mn2e}

\usepackage[usenames,dvipsnames]{color}
\usepackage{graphicx}
\usepackage{amsmath}
\usepackage{amssymb}
\usepackage{amsfonts}
\usepackage{units}
\usepackage{soul}
\usepackage{relsize}
\usepackage{url}
\usepackage{hyperref}
\usepackage{verbatim}
\usepackage{longtable}
\usepackage[T1]{fontenc}

\newcommand{\Hi}{H\,{\sc i}}
\newcommand{\Di}{D\,{\sc i}}
\newcommand{\Cii}{C\,{\sc ii}}
\newcommand{\Civ}{C\,{\sc iv}}
\newcommand{\Alii}{Al\,{\sc ii}}
\newcommand{\Siii}{Si\,{\sc ii}}
\newcommand{\Siiv}{Si\,{\sc iv}}
\newcommand{\Mgii}{Mg\,{\sc ii}}

\newcommand{\Oi}{O\,{\sc i}}
\newcommand{\omegab}{$\Omega_bh^2$} 					
\newcommand{\kms}{\, \mathrm{km\, s}^{-1} } 					
 					
\newcommand{\ang}{\, \AA{}} 					
\newcommand{\eV}{\, \mathrm{eV}} 					
\newcommand{\chidof}{$\chi^2/\mathrm{dof}$} 					

\newcommand{\pmol}{\, \mathrm{mol}^{-1} } 					
\newcommand{\popler}{\texttt{UVES\_popler}} 					
\newcommand{\vpfit}{\texttt{VPFIT}} 
\newcommand{\rdgen}{\texttt{RDGEN}} 
\newcommand{\cloudy}{\texttt{Cloudy}} 
\newcommand{\makee}{\texttt{MAKEE}} 
\newcommand{\Planck}{the {\sl Planck} Surveyor}

\newcommand\eqnref[1]{%
 Eqn.~\ref{eqn:#1}}

\newcommand\tabref[1]{%
Table~\ref{tab:#1}}

\newcommand\figref[1]{%
Fig.~\ref{fig:#1}}

\newcommand\secref[1]{%
Sec.~\ref{sec:#1}}

\newcommand\appref[1]{%
Appendix~\ref{app:#1}}

\usepackage[usenames]{xcolor}

\title[Robust deuterium measurement at z=3.256]{A robust deuterium abundance; Re-measurement of the $z=3.256$ absorption system towards the quasar PKS1937-1009}
\author[Riemer-S\o{}rensen et al.]{S. Riemer-S\o{}rensen$^{1,2,3}$\thanks{Email: signe.riemer-sorensen@astro.uio.no}, J. K. Webb$^{4}$, N. Crighton$^{5}$, V. Dumont$^{4}$, K. Ali$^{4}$ \newauthor 
 S. Kotu\v{s}$^{5}$,  M. Bainbridge$^{4}$, M. T. Murphy$^{5}$, R. Carswell$^{6}$ \\
$^{1}$Institute of Theoretical Astrophysics, The University of Oslo, Boks 1072 Blindern, NO-0316 Oslo, Norway\\
$^{2}$School of Mathematics and Physics, University of Queensland, Brisbane QLD 4072, Australia\\
$^{3}$ARC Centre of Excellence for All-sky Astrophysics (CAASTRO)\\
$^{4}$School of Physics, University of New South Wales, Sydney NSW 2052, Australia\\
$^{5}$Centre for Astrophysics \& Supercomputing, Swinburne University of Technology, P.O. Box 218, Hawthorn, VIC 3122, Australia\\
$^{6}$Institute of Astronomy, University of Cambridge, Madingley Road, Cambridge CB3 0HA, United Kingdom}

\begin{document}

\date{\today}
\pagerange{\pageref{firstpage}--\pageref{lastpage}} \pubyear{2014}

\maketitle

\begin{abstract}
The primordial deuterium abundance is an important tracer of the fundamental physics taking place during Big Bang Nucleosynthesis. It can be determined from absorption features along the line of sight to distant quasars. The quasar PKS1937-1009 contains two absorptions systems that have been used to measure the primordial deuterium abundance, the lower redshift one being at $z_{abs} = 3.256$. New observations of this absorber are of a substantially higher signal-to-noise and thus permit a significantly more robust estimate of the primordial deuterium abundance, leading to a \Di/\Hi{} ratio of $2.45\pm0.28\times10^{-5}$.  Whilst the precision of the new measurement presented here is below that obtained from the recent cosmological parameter measurements by \Planck, our analysis illustrates how a statistical sample obtained using similarly high spectral signal-to-noise can make deuterium a competitive and complementary cosmological parameter estimator and provide an explanation for the 
scatter seen between some existing deuterium measurements.

\end{abstract}

\section{Introduction}
The Standard Model of particle physics successfully describes the elementary particles and the forces binding them together, but it has significant problems with dark matter and neutrinos. The identity of the dark matter remains elusive and the neutrinos are described as exactly massless despite the observed neutrino oscillations which requires mass \citep{Fukuda:1998,Forero:2012,Beringer:2012}. Both problems can be solved by the existence of additional and so far unobserved particles \citep{Beringer:2012}. Such speculative particles may affect the formation of the light elements during the nucleosynthesis a few minutes after Big Bang \citep[e.g.][]{Steigman:2012, Nollett:2014, Boehm:2013,Archidiacono:2014}. This would be reflected in the light element abundances we observe today.
Precise measurements of the deuterium abundance combined with measurements of the helium abundance \citep{Izotov:2013} provide a constraint on the high value of relativistic degrees of freedom, $N_\mathrm{eff}$, preferred by current cosmological observations \citep{PlanckXVI:2013, Battye:2014, Wyman:2013, Riemer-Sorensen:2013} independently of the Cosmic Microwave Background (CMB), allowing us to distinguish between various hypothesised scenarios \citep{DiValentino:2013,Steigman:2013}. 

The deuterium abundance is an excellent ``baryometer'' sensitive to the number density of baryons, $n_{\mathrm{b},0}$, present in the early universe. This is usually quantified as the ratio between the number densities of baryons and photons 
\begin{equation}
\eta_{10}=10^{10}\frac{n_{\mathrm{b},0}}{n_{\gamma,0}} 
\approx 273.9\Omega_\mathrm{{b,0}}h^2 \, ,
\end{equation}
where the photon number density, $n_{\gamma,0}$, is determined from the CMB temperature \citep[e.g.][]{Fixsen:2009}, and $h$ is the unit-less Hubble parameter $H_0 = 100h\kms\mathrm{Mpc}^{-1}$.

The baryon density can also be measured from the CMB, with \Planck\ providing the very precise measurement of $\Omega_bh^2 = 0.02207\pm0.00027$
at the time of recombination \citep[{\sl Planck + WP + High $\ell$} analysis in][]{PlanckXVI:2013}.

The deuterium abundance can be determined from absorption features along the lines of sight to distant quasars. Previous measurements have shown an unnaturally large scatter \citep{Burles:1998b,Pettini:2001,Pettini:2008,Pettini:2012}, albeit with some recent improvement \citep{Cooke:2014}. The precision with which the nuclear reaction rates can be determined limits the constraints from nucleosynthesis, currently 2-4\% \citep{Burles:1999,Coc:2013}. \citet{Cooke:2014} states that their statistical uncertainty is of the order of 1\% while the systematic error from the nuclear rates is 2\%.

In this paper, which focuses on the  absorber at $z_\mathrm{abs} = 3.256$, we present a new and precise measurement of the deuterium abundance measured in the spectrum along the line of sight to PKS1937-1009 (1950) also known as J193957-100241 (J2000). It was identified as a quasar with emission redshift $z_\mathrm{em} = 3.787$ by \citet{Lanzetta:1991}. The presence of deuterium was previously identified in two independent absorption systems along the line of sight at $z_\mathrm{abs} = 3.256$ \citep{Crighton:2004} and $z_\mathrm{abs} = 3.572$ \citep{Burles:1998a}. Since those measurements were published, the quasar has been observed extensively with both the Very Large Telescope (VLT) and the Keck Telescope, effectively increasing the observation time by almost an order of magnitude. As a consequence of the dramatically increased signal-to-noise, the new data presented in this paper reveals a far more complex velocity structure in the absorbing gas -- \citet{Crighton:2004} detected only two components whereas we now detect five.

The observational data is presented in \secref{Observations}, followed by the details of our analysis in \secref{Analysis}, and results in \secref{Results}. The best fit model details are given in \appref{Model}. In \secref{Discussion} we compare our results to previous measurements and to the {\sl Planck} measurement of \omegab, and discuss the cosmological implications, before concluding in \secref{Conclusions}.

All quoted uncertainties are 68 per cent confidence level unless otherwise specified.

\section{Observations} \label{sec:Observations}
We have used observations from the archives of Ultraviolet and Visual Echelle Spectrograph (UVES) at VLT, and both the High Resolution Echelle Spectrometer (HIRES) and the Low Resolution Imaging Spectrometer (LRIS) at the Keck Observatory. The observations are listed in \tabref{data}.

\begin{table*}
\begin{minipage}{158mm}
\caption{The observation details}\label{tab:data}
{\centering
\begin{tabular}{@{}l l l l l l l@{}}
\hline
Date			& Primary 			& Instrument		& Setting				& Resolving 	& Resolution & Observation \\
			& investigator		& 				& (details)\footnotemark[1]& power 		& $\sigma_v [\kms]$\footnotemark[2]				& time [ks]			 \\ \hline
1996-08-09	& Songaila		& Keck LRIS		& w=0.7\arcsec			& 1500		& 400 			& 2.4	\\
1997-10-03	& Cowie			& Keck HIRES		& C5 (1.148\arcsec, 3910/6360\ang)	& 37000		& 3.5				& 4x2.4\\ 
1997-10-04	& Cowie			& Keck HIRES		& C5 (1.148\arcsec, 3910/6360\ang)	& 37000		& 3.5				& 2x2.4 + 1x1.4\\ 
2005-07-01	& Crighton		& Keck HIRES		& B5 (0.861\arcsec, 3630/8090\ang)	& 49000		& 3.3				& 6x3.6 \\ 
2005-08-12	& Tytler			& Keck HIRES		& C5 (1.148\arcsec, 3790/6730\ang)	& 37000		& 3.5				& 2x6.4 + 1x6.0\\ 
2006-04-10	& Carswell\footnotemark[3] & VLT UVES		& DICHR\#1 (1.0\arcsec, 3900/5800\ang)			& 45000		& 2.8 			& 1x5.4 \\
2006-06-01	& Carswell\footnotemark[3]		& VLT UVES		& DICHR\#1 (1.0\arcsec, 3900/5800\ang)			& 45000		& 2.8 			& 2x5.4 \\
2006-06-25	& Carswell\footnotemark[3]		& VLT UVES		& DICHR\#1 (1.0\arcsec, 3900/5800\ang)			& 45000		& 2.8 			& 1x5.4 \\
2006-07-21	& Carswell\footnotemark[3]		& VLT UVES		& DICHR\#1 (1.0\arcsec, 3900/5800\ang)			& 45000		& 2.8 			& 1x5.4 \\
\hline
\end{tabular}}
\footnotemark[1]{For Keck LRIS observations the slit width, for Keck HIRES observations the slit width, cross-disperser angle and central wavelength, and for VLT UVES the slit width and central wavelength of blue/red arm.}\\
\footnotemark[2]{The resolution in terms of one standard deviation spread in velocity space, $\sigma_v$, are estimated from the line width of a slit fully illuminated with a Thorium-Argon lamp. The actual value can be up to $~10\%$ higher. We have checked that there is no effect on $\chi^2/\mathrm{dof}$ from changing $\sigma_v$ by $\pm10\%$ in individual fit regions.}\\
\footnotemark[3]{All VLT UVES observations belong to the ESO programme with ID 077.A-0166(A).}
\end{minipage}
\end{table*}

\subsection{Keck LRIS observations} 
The LRIS provides a lower resolution spectrum where the Lyman limit is clearly visible and can be used to constrain the total column density of the system. It consists of a 40 minute exposure from August 1996 with a slit width of 0.7\arcsec \citep{Songaila:1997,Crighton:2004}. The spectrum was flux calibrated using a standard star observation to recover the quasar emission shape resulting in a coverage of the range $3700-7400$\ang\ with a signal to noise of 200 at $6000$\ang.

\subsection{Keck HIRES observations} \label{sec:Keck}
The Keck HIRES spectra were taken during three separate observing runs in 1997 and 2005 (see \tabref{data}). We obtained the exposures and calibrations for each run from the Keck Observatory Archive\footnote{\url{http://www2.keck.hawaii.edu/koa/public/koa.php}} and reduced them using the purpose written software package \makee\footnote{\url{http://www.astro.caltech.edu/~tb/makee/}}. For each exposure, \makee{} subtracts the bias and the sky background, corrects for the echelle blaze and extracts a one-dimensional spectrum for each echelle order. By identifying arc lines it finds a solution for each order and converts pixel number on the CCD to a wavelength scale. A custom-written Python code\footnote{Available for download at \url{https://github.com/nhmc/HIRES}} was used to combine the spectra from each order and exposure into a single spectrum. We estimate the unabsorbed continuum level by fitting spline curves to regions that are judged to be free from absorption. This is a straightforward procedure redward of the QSO Lyman $\alpha$ emission line, where many of the metal transitions fall. Bluewards of the QSO Lyman $\alpha$ emission however, where the hydrogen and deuterium transitions fall, it is more difficult to estimate the continuum due to line blanketing by the Lyman $\alpha$ forest. In section \secref{continuum} we discuss how we account for uncertainties in the continuum.

The individual spectra with similar settings and resolutions were stacked resulting in three final spectra together covering the range $4100-6400$\ang.

\subsection{VLT observations} \label{sec:VLT}
The VLT UVES observations were taken in April, June and July 2006 (see \tabref{data}). The one-dimensional reduction of the raw data and its calibration was completed using the Common Pipeline Language\footnote{\url{http://www.eso.org/sci/software/cpl/}} provided by ESO. The post-pipeline reduction process was done using the publicly available \popler\footnote{\url{http://astronomy.swin.edu.au/~mmurphy/UVES_popler/}} software including rebinning of all spectra to a common vacuum heliocentric wavelength scale, combining the echelle orders, removing leftover cosmic rays by automatic $\sigma$-clipping etc. \citep[procedure described in detail in][]{Bagdonaite:2014}. The automatic processing was followed by a visual inspection of pixel values with large discrepancies between the individual orders or exposures and removal of suspicious values, in particular at the edges of each order.

Since the three VLT observations were taken with identical settings, all exposures were stacked resulting in a single spectrum for further analysis. The final spectrum covers the wavelength range 3821 to 6813\ang{} with a number of gaps\footnote{The gaps are at 4520--4781\ang, 4809--4812\ang, 4847--4850\ang, 4923--4928\ang, 5760--5836\ang, 6065--6069\ang.} and an average signal to noise of 57 around Lyman $\alpha$ and 26 around Lyman $\beta$.

In order to normalise the spectrum, the continuum was fitted during the post-pipeline processing. The spectrum was divided into overlapping chunks of $15000\kms$ in the Lyman forest and $2000\kms$ in the red part, and each chunk was fitted with a low order Chebyshev polynomial (4th order in the forest, and 6th-8th order elsewhere). The chunk fits were subsequently spliced to obtain a smoothly varying low order Chebyshev polynomial rather than a high order polynomial from fitting the entire spectrum simultaneously with a single function.
The continuum over the absorption regions was determined by connecting absorption-free regions in the neighbourhood. Since the main aim is to fit a continuum over the entire observed quasar emission and leave out the absorption, the continuum fit had different threshold filters for higher and lower flux values ($3\sigma$ above, and $1.2\sigma$ below in the forest, and $1.4\sigma$ below in the red part). The harsher lower limit reduces the influence of the absorption regions. 

There is a problem with the spectral error estimation in the UVES pipeline,\footnote{Described in Carswell, R.F. \rdgen{} \url{http://www.ast.cam.ac.uk/~rfc/rdgen9.5.pdf} as well as in \citet{King:2012}.} where the accuracy of the uncertainty estimation is significantly reduced at the base of saturated absorption lines compared to the scatter of the individual measurements. We have followed the steps outlined in \citet[][section 2.1]{King:2012} to rectify the issue.

\section{Analysis and results} \label{sec:Analysis}
\subsection{Voigt Profiles}
The relevant absorption lines are fitted with Voigt profiles \citep{Hjerting:1938,Carswell:1984} using the software package \vpfit\footnote{Version 10, Carswell, R.F. and Webb, J.K. \vpfit{} - Voigt profile fitting program \url{http://www.ast.cam.ac.uk/~rfc/vpfit.html}}. It is based on a $\chi^2$ minimisation process to find the best-fitting redshift (z), column density ($\log N$ in cm$^{-2}$) and dynamic line broadening ($b$ in $\kms$) for each sub-component of the absorption system. The dynamic broadening is a combination of turbulent motion in the gas and thermal broadening given by
\begin{equation} \label{eqn:bparam}
b =2\sigma^2 = b_\mathrm{turb}^2+2kT/m \, ,
\end{equation}
where $\sigma$ is the Gaussian velocity dispersion of the gas, $k$ is Boltzmann's constant, $T$ is temperature, and $m$ is mass of the atomic species. \vpfit{} gives the option of relating the $b$-parameters of different species under the assumption of purely thermal broadening (the gas temperature determines the broadening), purely turbulent broadening (the gas temperature contribution is negligible), and if several species are available both the thermal and turbulent terms can be evaluated. 

\subsection{Fitting}
\begin{figure*}
\centering
\includegraphics[width=0.99\textwidth]{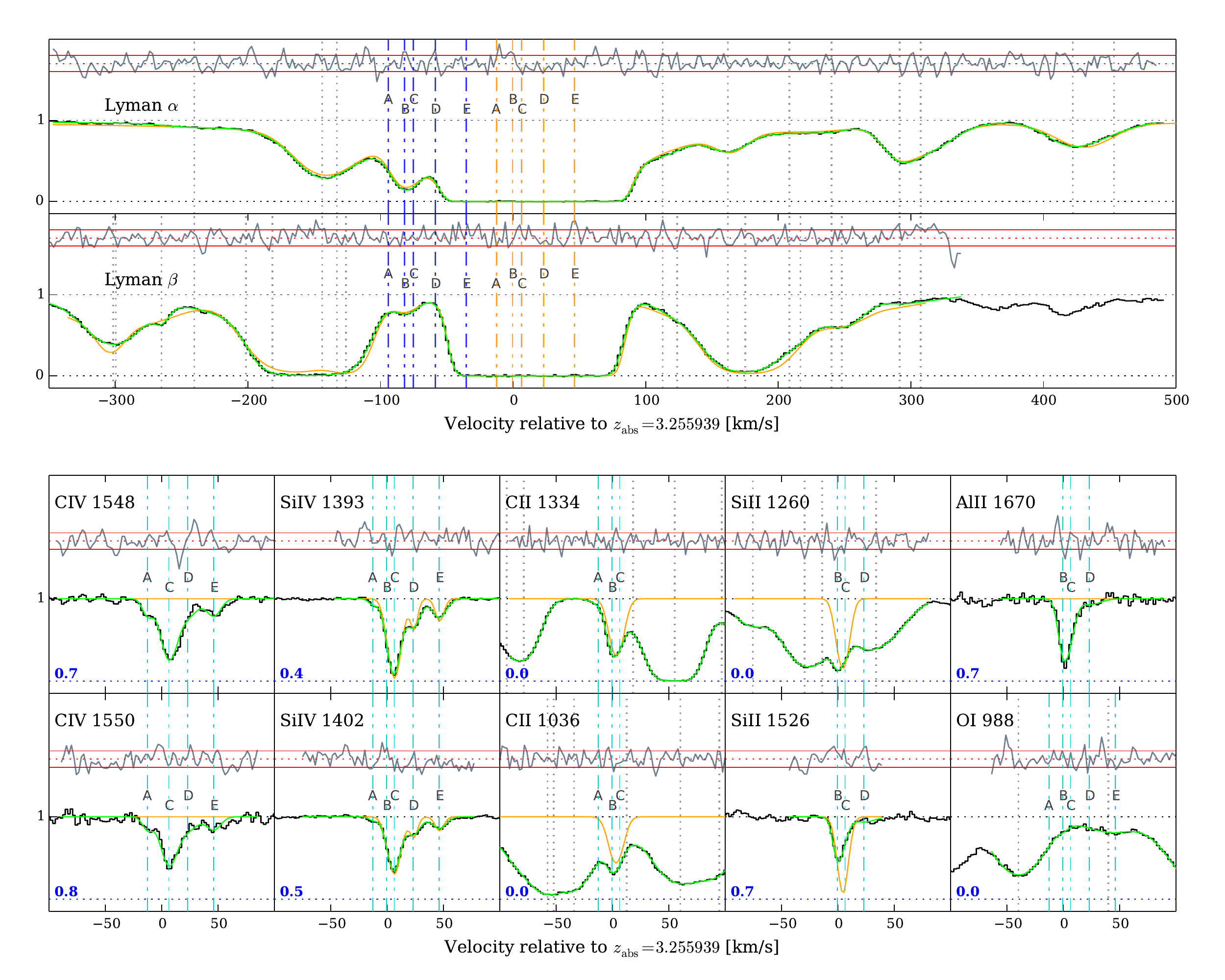}
\caption{The stacked spectra (weighted by uncertainty, black) and the best fitting model (green/light grey) as well as the normalised fit residuals (above each fit). The composite spectrum is used only for visualisation purposes as the individual data sets are always fitted simultaneously rather than stacked. The velocity components A to F are marked by vertical dot-dashed lines (\Hi{} in orange/light grey, \Di{} in blue/dark grey, metals in red/grey) and interloping \Hi{} lines in vertical dotted lines (light grey). The solid orange line shows the old model from \citet{Crighton:2004} discussed in \secref{previous}.}
\label{fig:spectrum}
\end{figure*}

The spectra described in \secref{Observations} are kept as independent spectra, rather than combined into one composite spectrum, and the whole set are fitted simultaneously. In total we have 10 metal regions in the system (\Civ{} 1548, \Civ{} 1550, \Siiv{} 1393, \Siiv{} 1402, \Cii{} 1334, \Cii{} 1036, \Siii{} 1260, \Siii{} 1526, \Alii{} 1670, and a \Mgii{} 2803 blend at $z=0.560207$) and two Lyman regions available (Lyman $\alpha$ and $\beta$). For visualisation we use a weighed stacked spectrum as shown in \figref{spectrum}. Since we only have coverage of the first two Lyman series lines in this absorption system, the hydrogen $b-$parameter and column density are not independently well-determined. We explore the consequences of this in Sec. \ref{mod:temp}, \ref{mod:turb}, \ref{mod:alluntied}.

For all species we vary the summed column densities rather than the individual column densities so as to derive a more accurate estimate of the {\it total} column density across the absorption complex. This is achieved by assigning one of the $m$ column density parameters to be the column density summed over all $m$ components, leaving $m-1$ column density parameters assigned to the remaining individual values. This procedure is beneficial in these circumstances because the summed column density may be well-constrained, whereas the column densities for individual velocity components may not be. We require the ratio of $N$(\Di{})/$N$(\Hi{}) to be identical in all subsystems (velocity components). This equates to assuming the same degree of astration for all components.  The overall metalicity of this system is sufficiently low for this to be a reasonable assumption (\secref{metallicity}). To check for systematic deviations between the telescopes, we fit each model to the Keck spectra alone, the VLT spectra alone, and to the entire data set including the LRIS data as well.

\subsubsection{Continuum and velocity shifts}\label{sec:continuum}
The spectra are normalised to the quasar emission continuum as part of the initial data processing. There it is assumed that the continuum shape does not change abruptly over small ranges but only varies over long ranges. To allow for any possible local variations, we add a flat continuum to the Voigt profile model around all Lyman lines (and some of the metal lines) to account for any uncertainty in the initial continuum modelling. If the continua slopes are treated as free parameters, the preferred fit values are consistent with zero slope and the the improvement in $\chi^2$ per degree of freedom (\chidof) is less than 0.01. 

We also allow for a velocity shift between the observations to account for any calibration or slit-centering offsets. The shifts are required to be identical for all regions from the same telescope exposure. The only exception is one of the \Siiv{} regions observed with Keck. The preferred velocity shift differed to the other regions in the same exposure so we introduced that additional free parameter in the fit. We checked that the \chidof{} did not improve by allowing any of the other regions to shift independently. The best fit velocity shifts (given in \tabref{model}1) are applied before all spectra are combined for visualisation in \figref{spectrum}.

\subsection{Models}\label{sec:models}
The composite spectrum in \figref{spectrum} clearly shows an absorption line with an off-set of $~85.2\pm8.7\kms$ from the centre of the main hydrogen component in both Lyman $\alpha$ and $\beta$. This fits well with the expected off-set of $85\kms$ and the lines are clearly narrower than the nearby hydrogen lines and wider than expected for thermally broadened metal lines. These facts strongly suggest that the feature is indeed from \Di{}. It is apparent from the spectrum that \Siiv{} has at least three velocity components, whereas \Siii\ has a slightly different structure not entirely aligned with that of \Siiv{}. This is not unexpected as \Siii{} and \Siiv{} have very different ionisation potentials (see \tabref{ions}) and are known not always to trace the velocity structure of the lower ionisation species \citep{Wolfe:2000,Fox:2007}. 
Consequently we group species according to their ionisation potentials and tie line fitting parameters accordingly. We then expect to find the same redshift components in \Civ{} and \Siiv{}, which may not correspond to those seen in \Cii{}, \Siii{}, and \Alii{}. All components are linked to a \Hi{} component with corresponding \Di{}. 

\begin{table}
\caption{Atomic properties}\label{tab:ions}
\begin{tabular}{@{}l r r r r}
\hline
Element		& Atomic mass		& $\Delta E_{1}	$	& $\Delta E_{2}	$		& $\Delta E_{4}$ \\ \hline
H			& 1.01			& 13.60			& ---				& --- \\
D			& 2.01			& 14.51			& ---				& ---\\
O			& 16.00			& 13.62 			& 35.12 			& 54.94 \\
C			& 12.01			& 11.27 			& 24.39			& 64.50 \\
Al			& 26.98			& 6.00			& 18.83			& 119.99 \\
Si			& 28.09			& 8.16			& 16.34			& 45.14	\\

\hline
\end{tabular}
\\
{The atomic masses are in units of $\mathrm{g}\pmol$. $\Delta E_{i}$ is the ionisation energy for the $i$th ionisation level in units of $\eV$.}
\end{table}

The \Siiv{}--\Civ{} complex is best fitted by five components (A, C, D, E, F in \figref{spectrum}), for which the column density of  \Civ{}(B) is too low to affect the fit ($\log(N)<8$). \
An automated model-building software based on genetic algorithms preferred six components, when fitted to \Siiv{}-\Civ{} alone, but the additional component was lost when tying it to \Hi{}. 

The \Cii{}--\Siii{}--\Alii{} complex is fitted by two of the \Siiv{}--\Civ{} components (B, C) as well as one additional component for \Cii{} (A) and one for \Siii{}--\Alii{} (D) both sharing redshifts with the \Siiv{}--\Civ{} complex. \Oi{} is blended with forest absorption, but we add it to the model tied to the low ionisation group in order to constrain the metallicity (see \secref{metallicity}).

\subsubsection{Temperature fitting} \label{mod:temp}
Most of the components have multiple metals allowing us to fit simultaneously for both temperature and $b$-parameters (see \eqnref{bparam}). This was done for the lower ionisation group, \Cii, \Siii, \Alii, \Hi, \Di.
The degeneracy between the two parameters depends on the number of available species and their mass differences. By simultaneously solving for temperature we ensure the resulting \Di/\Hi{} estimate and its 
error estimate correctly reflect this additional uncertainty in the model. The fitted temperature values and associated error estimates are presented in \tabref{model}1.

The ionisation potentials of \Siiv{} and \Civ{} are very different from those of \Cii{}, \Siii{}, \Alii{}, \Hi{}, \Di{}, and consequently the temperatures for the two groups of gasses should be allowed to vary independently.
However, the mass difference between \Siiv{} and \Civ{} is too small to recover both temperature and the turbulent $b$-parameter without additional information, so their $b$-parameters were kept tied. To decide whether the tying should be thermal or turbulent we fitted the entire model with the $b$-parameters un-tied (one fit for each of the components) and compared the preferred values to the ones expected for pure thermal or turbulent broadening. Component A, C, and D were clearly closer to thermal, while E was closer to turbulent. There were no significant differences in the \chidof\ between the models with free parameters and the ones with thermal and turbulent ties, but the tying reduces the number of free parameters and consequently the uncertainty of the \Di/\Hi{} ratio. However, the stronger physical assumptions about the gas structure may lead to increased systematic uncertainty. Component E was tied turbulently between \Siiv{}, \Civ{}, \Hi{} and \Di{} based on a similar reasoning.

This is our baseline model with a best fit \chidof\ $=1.13$ (see \figref{spectrum}), and unless otherwise stated the one for which we quote our main results (see \appref{Model} for the entire model).

\subsubsection{Tying everything turbulently} \label{mod:turb}
Based on $\chi^2$ comparison, \citet{King:2012} found that approximately 70 per cent of all absorption systems are better fitted by a model with turbulently tied $b$-parameters than one with thermal ties. We examined the $b$-parameters from the baseline model and performed a fit with the $b$-parameters of all subsystems tied turbulently within the ionisation groups. The fit is slightly worse than for temperature fitting (\ref{mod:temp}) with \chidof\ $=1.16$ instead of \chidof\ $=1.13$, but the resulting \Di/\Hi{} ratio of $2.46\pm0.11 \times 10^{-5}$ is completely consistent with the result of the temperature fit of $2.45\pm0.28 \times 10^{-5}$. The reduced number of free parameters leads to very small uncertainties on \Di/\Hi{}, but at the price of strong assumptions about the gas. Although the majority of systems can be approximated by a turbulent fit, the line broadening is a combined effect of thermal and turbulent broadening and when possible one should fit for both.

\subsubsection{All untied} \label{mod:alluntied}
To check that the imposed assumptions on the $b$-parameters are reasonable, we also fitted the baseline model with all $b$-parameters and temperatures untied, allowing each component to have different broadening for each species regardless of which ionisation group it belongs to. However all redshift components were kept tied as for the other models listed, which implies inconsistent physical assumptions - untying \emph{b}-parameters infers the different species are no longer considered co-spatial, yet keeping redshifts tied maintains the co-spatial assumption. Nonetheless, by obliging \vpfit\ to maintain the same velocity structure, we get a useful indication of how the final uncertainty estimate on \Di/\Hi{} in the preferred model benefit from tying the \emph{b}-parameters in a physically plausible way.

The untying of the $b$-parameters results in a slight decrease of \chidof\ (from 1.13 to 1.12 for 5562 and 5756 degrees of freedom respectively) and a doubling of the \Di/\Hi{} uncertainty, but with the central values consistent ($2.45\pm0.28\times10^{-5}$ and $2.50\pm0.59\times10^{-5}$). The small change in \chidof\ indicates that the baseline model describes the data well, and that little is gained by introducing the extra degrees of freedom by relaxing the physical assumptions of tied temperatures and turbulent flow.

\subsubsection{No \Siiv\ and \Civ} \label{mod:simple}
Due to the large difference in ionisation potential between \Siiv-\Civ and the lower ions, the \Siiv-\Civ\ absorbers do not necessarily contain significant amounts of \Hi{}. To check the effect of this assumption we have fitted the base model (\ref{mod:temp}) excluding the \Siiv-\Civ\ regions (varying the continuum normalisations and slopes over Lyman $\alpha$ and $\beta$ freely). This results in \Di/\Hi{} of $2.40\pm0.30\times10^{-5}$ with a \chidof =$1.09$ for $4609$ degrees of freedom consistent with the base model result.

\subsection{Deuterium and hydrogen results}\label{sec:Results}
The deuterium and hydrogen column densities and their ratios from the various models are presented in \tabref{results} with the main result being a \Di/\Hi{} ratio of $2.45\pm0.28\times10^{-5}$ from the temperature fit (Sec. \ref{mod:temp}). We consider the temperature fit as providing the most realistic parameter error estimates. As \tabref{results} illustrates, the turbulent model yields a \Di/\Hi{} error estimate three times smaller, but this error is implausibly small and may be a consequence of the broadening assumptions made and the artificially reduced number of free parameters.

We interpret the consistency across the models as an indication of robust modelling of the absorption system in the sense that relaxing the various assumptions does not alter the final result by more than $1\sigma$, but affects the uncertainties as it allows for more freedom in the models. Similarly there is agreement between the analyses of the spectra from the two different telescopes to within $1\sigma$. 

The quoted uncertainties are purely statistical, but the agreement between the different models and the different data sets suggest a robust measurement.

\begin{table*}
\begin{minipage}{168mm}
\caption{Results of the models described in \secref{models}} \label{tab:results}
{\centering
\begin{tabular}{@{}l l c c c c c@{}}
\hline
Model							& Spectra		& $\chi^2$	& $\log(N$(\Di))			& $\log(N$(\Hi))		& $\log(\mathrm{D/H})$		& \Di/\Hi{} [$\times 10^{-5}$] \\ \hline
{\bf Temperature fit (\ref{mod:temp})}		& {\bf All}		& {\bf 1.13}	& {$\bf 13.48\pm0.0433$} & $\boldsymbol{18.09\pm0.0346}$	& $\boldsymbol{-4.610\pm0.0502}$	& $\boldsymbol{2.45 \pm 0.28}$\\[5pt]
Temperature fit (\ref{mod:temp})		& Keck+LRIS	& 1.13		& $13.48\pm0.0643$	& $18.08\pm0.0521$	& $-4.602\pm0.0828$			& $2.50\pm0.48$\\[5pt]
Temperature fit (\ref{mod:temp})		& VLT+LRIS	& 1.02		& $13.46\pm0.0586$	& $18.11\pm0.0470$	& $-4.647\pm0.0751$			& $2.26\pm0.39$\\[5pt]
Turbulent ties (\ref{mod:turb})		& All                & 1.16		& $13.46\pm0.0110$	& $18.07\pm0.0168$	& $-4.609\pm0.0201$			& $2.46\pm0.11$\\[5pt]
All untied (\ref{mod:alluntied})		& All		& 1.12		& $13.51\pm0.0566$	& $18.10\pm0.0813$      & $-4.590\pm0.0991$			& $2.57\pm0.59$\\[5pt]
No \Siiv-\Civ\ (\ref{mod:simple})		& All		& 1.09		& $13.40\pm0.0188$	& $18.02\pm0.0519$      & $-4.620\pm0.0552$			& $2.40\pm0.30$\\

\hline
\end{tabular}}
We have kept three significant figures on the logarithmic parameter errors here merely to illustrate the origin of the final error on our (non-logarithmic) \Di/\Hi{} ratios in the last column. 
\end{minipage}
\end{table*}

\subsection{Metallicity} \label{sec:metallicity}
\citet{Fields:2001} have explored several chemical evolution models, the simplest of which suggests that for gas with metallicities below 0.1 solar, very little deuterium depletion could have occurred and hence such probes are likely to yield a primordial \Di/\Hi{} measurement. The same is not true for bimodal star formation models.

The Voigt profile modelling of the observational data yields measurements of the \Hi, \Oi, \Cii, \Alii, \Siii{}, \Civ{} and \Siiv{} column densities in individual velocity components across the absorption system. The best-fit parameter results are given in \tabref{model}1 (in \appref{Model}). 

We assumed the gas to be in photoionisation equilibrium with the ambient ultraviolet background and used \cloudy{} \citep{Ferland:2013} to generate gas cloud models whose parameters span an appropriate range for this absorption system. We selected a Haardt-Madau HM05 ultraviolet background\footnote{See Section 6.11.8 in \url{http://www.nublado.org/browser/branches/c13_branch/docs/hazy1_c13.pdf}} to calculate \cloudy{} models with plane-parallel geometry. 
Since neither the neutral hydrogen volume density, $n_\mathrm{H}$, nor metallicity, $Z$, are known {\it a priori}, the initial parameter ranges are kept broad and we computed models in a coarse grid across the particle density range $-5 < \log n_\mathrm{H}[cm^{-3}] < 3$.

\begin{figure}
\centering
\includegraphics[width=0.99\columnwidth]{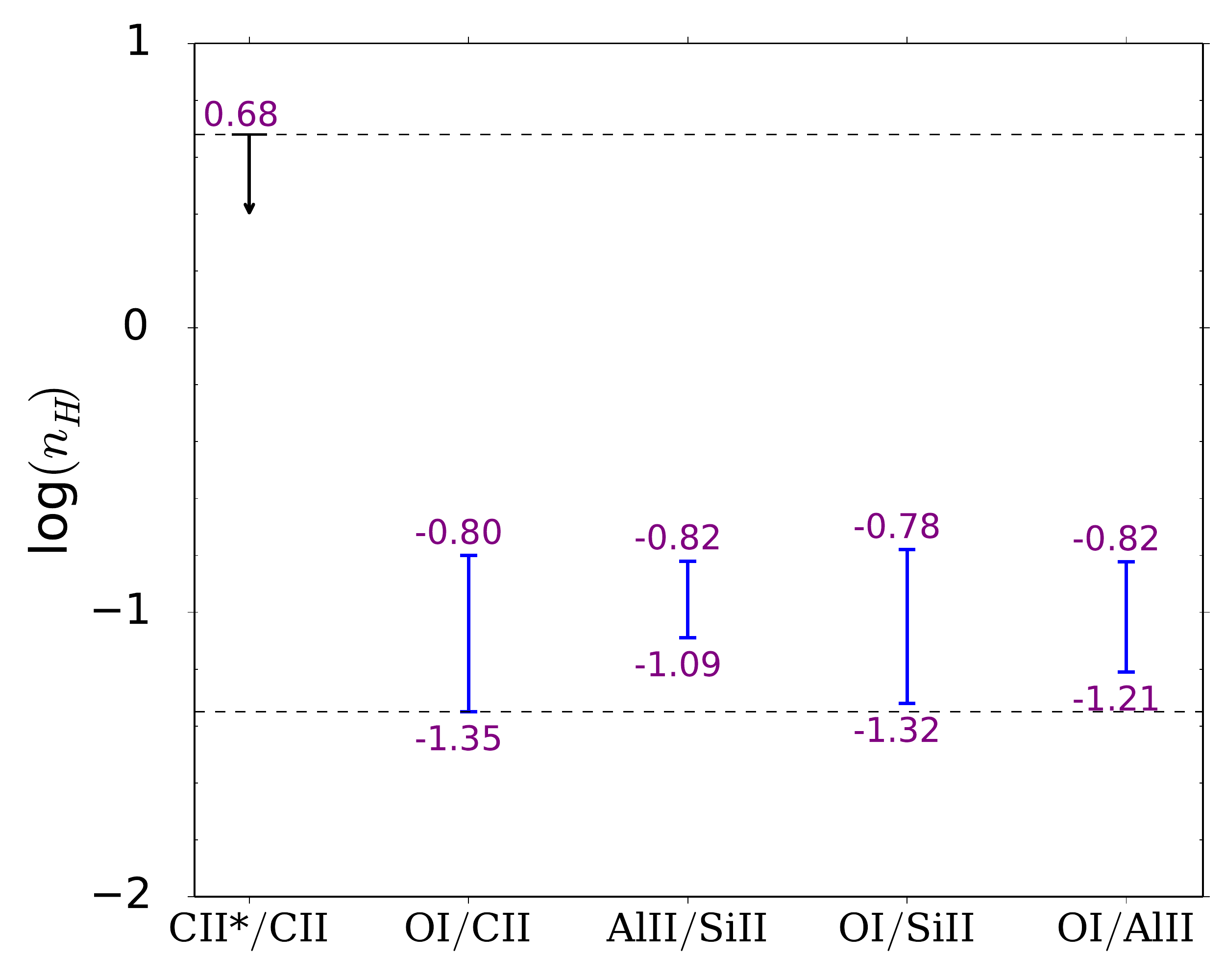}
\caption{Inferred particle density range from \cloudy{} models based on the observational (\vpfit) column density ratios. The upper limit from \Cii*/\Cii{} is calculated by using Eqn. 61 from \citet{Bahcall:1968} and assumes fully neutral hydrogen. In practice, there will be partial ionisation and the resulting upper limit will have a smaller value (see text) which in good agreement with our $\log n_\mathrm{H}$ range. The ratio \Siii/\Cii{} has been ignored since it provides two constraints, one which agrees with other ionisation species over a larger range and another which puts a lower limit of $\log n_\mathrm{H} [cm^{-3}]> 2.25$ disagreeing with \citet{Bahcall:1968} derived limits.}
\label{fig:nH}
\end{figure}

\Oi/\Hi{} and \Siii/\Hi{} are relatively insensitive to $n_\mathrm{H}$, so a comparison between \cloudy models and the observed values immediately provides a constraint of approximately $-2.6 \lesssim \log(Z/Z_{\sun}) \lesssim -1.1$. Over this metallicity range, \cloudy{} models indicate very little sensitivity of the predicted \Alii/\Siii{} column density ratio to $Z$, but strong sensitivity to $n_\mathrm{H}$. Another important ratio is \Oi/\Siii{} since these two species are the least susceptible to depletion effects given our system properties. Comparing \cloudy{} and observed \Oi/\Siii{}, we derive an upper limit on the particle density. The lower limit was computed by considering the \Oi/\Cii{} ratio which gives a conservative range on the allowed $n_\mathrm{H}$ range. \figref{nH} shows the determined $\log n_\mathrm{H}$ ranges using several pairs of species. As can be seen, the ionisation pairs agree quite well with one another which leads to the resulting particle density constraint $-1.35 < \log n_\mathrm{H} [cm^{-3}]< -0.78$.

We can compare the range in $\log n_\mathrm{H}$ derived using \cloudy{} models with that derived using the \Cii*/\Cii{} ratio, following the method of {\it detailed balancing} using Eqn. 61 in \citet{Bahcall:1968}. We have ignored the proton collision rate since this should be negligible at the prevailing gas temperature. The electron and hydrogen collision rates were obtained using Eqn. 2 in \citet{Blum:1992} and interpolating the values in Tab. 2 of \citet{Keenan:1986} respectively. \Cii* is not detected in the system but its absence was used to derive an upper limit on the column density. We imposed parameter constraints by assuming the redshifts and $b$-parameters of \Cii{} and corresponding putative \Cii* absorption components to be the same. We required N(\Cii*) to scale by a common factor to N(\Cii) over all corresponding velocity components i.~e. N(\Cii*) = f N(\Cii). \vpfit{} was then used in an iterative procedure to compute a series of models, for $f > 0$, increasing the value of $f$ until the best-fit $\chi^2_\mathrm{min}$ increased by unity (for $1\sigma$, given only one parameter being varied). This procedure yielded independent constraints, $\log n_\mathrm{H}[cm^{-3}]< 0.68$ and $\log n_e [cm^{-3}]< -0.87$, where $n_e$ is the electron density, which are both consistent with, but considerably less precise than the values derived from the \cloudy{} models.

Our constraint of $-1.35 < \log n_\mathrm{H}[cm^{-3}]< -0.78$ may be compared with hydrodynamical simulations of the Lyman-$\alpha$ forest; \citet{Dave:1998} provide a volume density--column density relation which would suggest a particle density around $ \log n_\mathrm{H} [cm^{-3}]= -2$ in our case, inconsistent with our observations. The higher ionisation species \Civ{} and \Siiv{} tend to favour a lower particle density environment and this discrepancy could be resolved by considering single component \cloudy{} models where these species were detected.

A further consistency check can be derived by comparing the gas temperature returned by the \cloudy{} models with the temperature estimated during the \vpfit ing process by solving simultaneously for the turbulent component of the given b-parameter and temperature, using several species. 
The temperature is estimated as a free parameter for each velocity component in the absorption system. One of the velocity components (B in \figref{spectrum}) which absorbs in all of \Hi, \Cii, \Siii\ and \Alii{} yields a reasonably small error estimate for the gas temperature, $T_{VPFIT} = 15000 \pm 800$K (see \appref{Model}). The other four components have only poorly determined \vpfit\ temperatures. The \cloudy\ models relate to the average properties over all five components and gives a temperature in the range $T_\mathrm{\cloudy{}} = 11,205 - 12,750$K. The apparent \vpfit /\cloudy{} temperature difference is thus not unexpected.

\begin{figure}
\centering
\includegraphics[width=0.99\columnwidth]{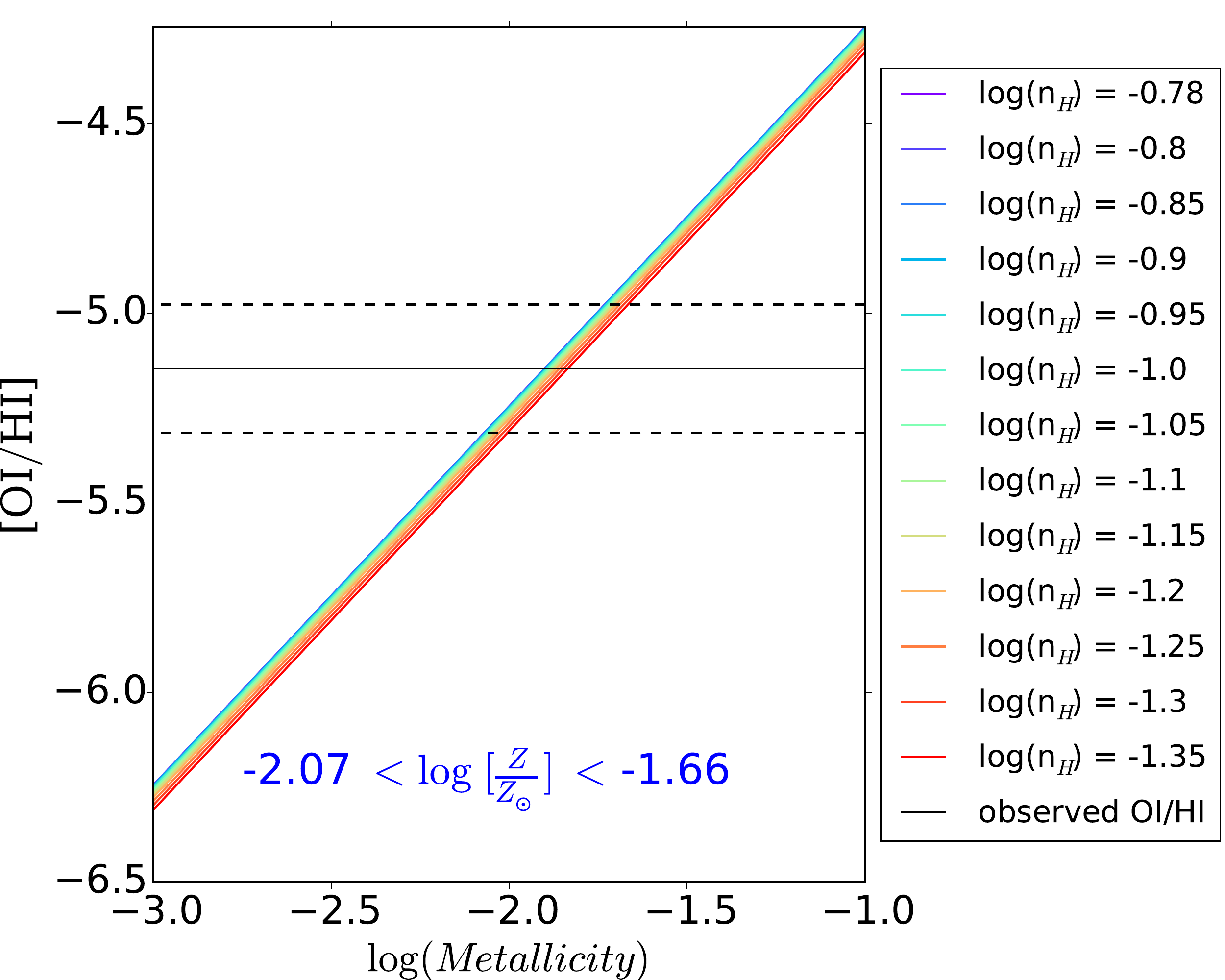}
\caption{\Oi/\Hi{} column density ratio as a function of metallicity for increasing particle densities (from upper to lower, range adopted from \figref{nH}). The observational column density is used to infer the allowed metallicity range with our \cloudy{} model.}
\label{fig:OIHI}
\end{figure}

In contrast to \Alii/\Siii{}, the \cloudy{} model \Oi/\Hi{} ratio is sensitive to $Z$ but insensitive to $n_H$, over the particle density range relevant here. This can be seen from \figref{OIHI} which indicates an increased sensitivity at lower particle density, however, the constraints developed above enable us to derive a metallicity range of $-2.07 < \log(Z/Z_{\sun}) < -1.66$.

\section{Discussion} \label{sec:Discussion}
\subsection{Line blending}
We searched the spectrum for any heavy element absorption systems with commonly found metal lines (Al, Fe, Mg, Si, C) and found tentative systems at $z=[$0.56030, 0.56790, 0.60312, 0.89683, 3.00851, 3.09545, 3.25618, 3.29134, 3.45359, 3.55357, 3.57230]. For each of the detected systems we checked whether any metal lines fall in the fitting regions at $z = 3.256$, i.e. the system we used to measure \Di/\Hi.
One blend was identified: \Mgii{} at $z = 0.5603$ falls near \Hi{} 1025 in our primary system. To properly account for this we fitted the \Mgii{} doublet simultaneously in our base model.

\subsection{Keck versus VLT}
The results from fitting the data from each telescope separately are consistent within $1\sigma$ albeit the VLT data prefers a lower value of (\Di/\Hi)$_\mathrm{VLT} = 2.26\pm0.39\times10^{-5}$ than the Keck data (\Di/\Hi)$_\mathrm{Keck} = 2.50\pm0.48\times10^{-5}$. Given the difference is well within the statistical uncertainty, the data give us no reason to suspect any kind of systematic explanation, so we ascribe the
difference to a random fluctuation.

\subsection{Comparison to previous measurements} \label{sec:previous}
\citet{Crighton:2004} measured the deuterium abundance in the same absorption system to be \Di/\Hi=$1.6^{+0.25}_{-0.30}\times 10^{-5}$ using a smaller sample of Keck data. This value is $3\sigma$ below the value obtained here. As shown in \figref{spectrum} (solid orange) the models appear visually quite similar but there are clear differences. The different \Di/\Hi{} values are explained by a combination of velocity structure differences and fitting procedures. Compared with \citet{Crighton:2004} we increase the observation time by an order of magnitude\footnote{Note we do not include the same data as they used in order to keep the two measurements completely independent.} 
The lower signal-to-noise data revealed fewer velocity components and \citet{Crighton:2004} assumed purely thermal broadening between \Hi{} and \Di{}, while we have more components and fit for both thermal and turbulent broadening. The re-analysis we present here therefore highlights the importance of high signal-to-noise in deriving a reasonable estimate of the velocity structure in quasar absorption system modelling.  The new measurement is closer to the CMB inferred value, but that is not an argument for robustness. Instead the increase in data quality combined with our tests of model assumptions points towards a more robust measurement. 

We compare our measurement to the sample from \citet{Pettini:2008} plus newer low uncertainty measurements from \citet{Fumagalli:2011}, \citet{Noterdaeme:2012} and \citet{Cooke:2014} listed in \tabref{measurements}. While this is by no means a complete sample, it serves to illustrate the issues and potential of future measurements. 

\begin{figure*}
\centering
\includegraphics[width=0.99\textwidth]{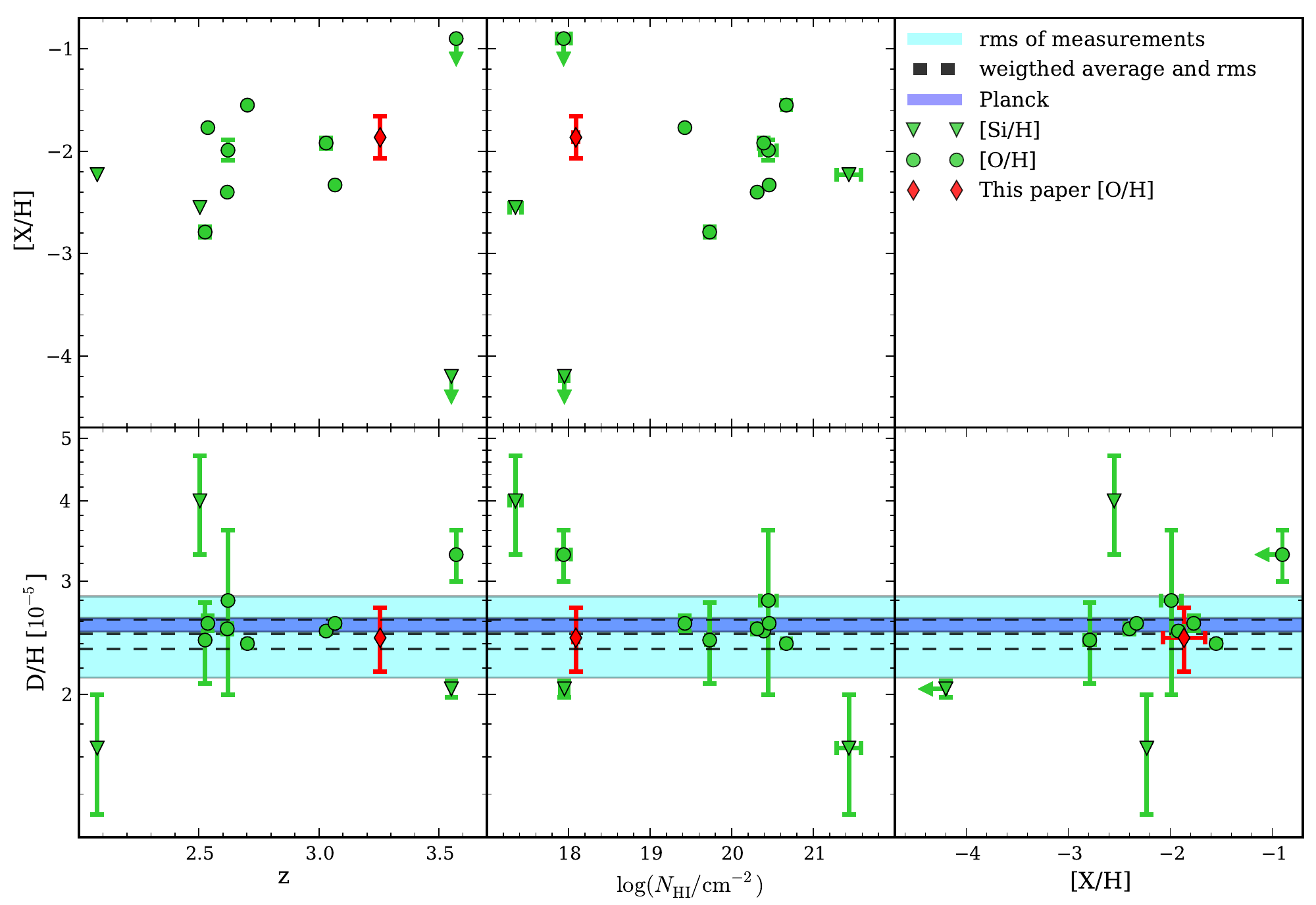}
\caption{Correlations between measured deuterium abundance and redshift (z), \Hi{} column density ($\log N$(\Hi)), and metallicity ([X/H]) for the sample defined in \tabref{measurements}. The uncertainties are plotted where given in the literature. The shaded horizontal bands show the {\sl Planck} constraint from standard BBN (inner), the weighted average of the entire sample (dashed, $2.48\pm0.13\times10^{-5}$) and the variance of the sample (outer).}
\label{fig:correlations}
\end{figure*}

\figref{correlations} shows a comparison of the absorption system properties of the measurements given in \tabref{measurements}. There is no apparent correlation between metallicity, redshift, deuterium or hydrogen column density (rank correlation tests show that the weak anti-correlation between deuterium and column density is not significant).

\begin{table*}
\begin{minipage}{150mm}
{\centering
\caption{A sample of robust measurements defined by \citet{Pettini:2008} plus newer low uncertainty measurements}\label{tab:measurements}
\begin{tabular}{@{}l c c c c c@{}}
\hline
Reference						& Absorption	& $\log N($\Hi{})             & [X/H]            & \Di/\Hi{}                    & 100\omegab \\ 
								& redshift	&                                      &                     & $[\times 10^{-5}]$       &                              \\ \hline
\citet{Burles:1998a}					& 3.572		& $17.9\pm0.08$            & <0.9 O         & $3.30\pm0.30$           & $1.86\pm0.11$    \\
\citet{Burles:1998b}				& 2.504		& $17.4\pm0.07$            & -2.55 Si       & $4.00\pm0.70$           & $1.65\pm0.18$    \\
\citet{Pettini:2001}					& 2.076		& $21.4\pm0.15$            & -2.23 Si       & $1.65\pm0.35$           & $2.91\pm0.52$    \\
\citet{Kirkman:2003}				& 2.426		& $19.7\pm0.04$            & -2.79 O       & $2.43\pm0.35$           & $2.26\pm0.20$    \\
\citet{Fumagalli:2011}				& 3.552		& $18.0\pm0.05$            & -4.20 Si       & $2.04\pm0.06$           & $2.52\pm0.05$    \\
\citet{Noterdaeme:2012}				& 2.621		& $20.5\pm0.10$		& -1.99 O       & $2.80\pm0.80$           & $2.07\pm0.37$    \\	
\citet{Cooke:2014}, \citet{Pettini:2012}	& 3.030		& $20.4\pm0.003$		& -1.92 O       & $2.51\pm0.02$           & $2.21\pm0.02$    \\
\citet{Cooke:2014}, \citet{OMeara:2001} & 2.537		& $19.4\pm0.01$		& -1.77 O       & $2.58\pm0.07$           & $2.18\pm0.04$    \\
\citet{Cooke:2014}, \citet{Pettini:2008}	& 2.618		& $20.3\pm0.01$		& -2.40 O       & $2.53\pm0.05$           & $2.20\pm0.03$    \\
\citet{Cooke:2014}					& 3.067		& $20.5\pm0.01$		& -2.33 O       & $2.58\pm0.03$           & $2.18\pm0.02$    \\
\citet{Cooke:2014}, \citet{OMeara:2006} & 2.702		& $20.7\pm0.05$		& -1.55 O       & $2.40\pm0.03$           & $2.20\pm0.03$    \\
This work						& 3.255		& $18.1\pm0.03$		& -1.87 O			& $2.45\pm0.28$       & $2.24\pm0.16$      \\	
\hline
\citet{PlanckXVI:2013}\footnotemark[1]	& ---			& ---					& --- 					& ---					& $2.207 \pm 0.027$\\ \hline
\end{tabular}}
\footnotemark[1]{Standard flat $\Lambda$CDM cosmology for {\sl Planck+WP+High $\ell$}}\\
\end{minipage}
\end{table*}

\begin{figure}
\centering
\includegraphics[width=0.99\columnwidth]{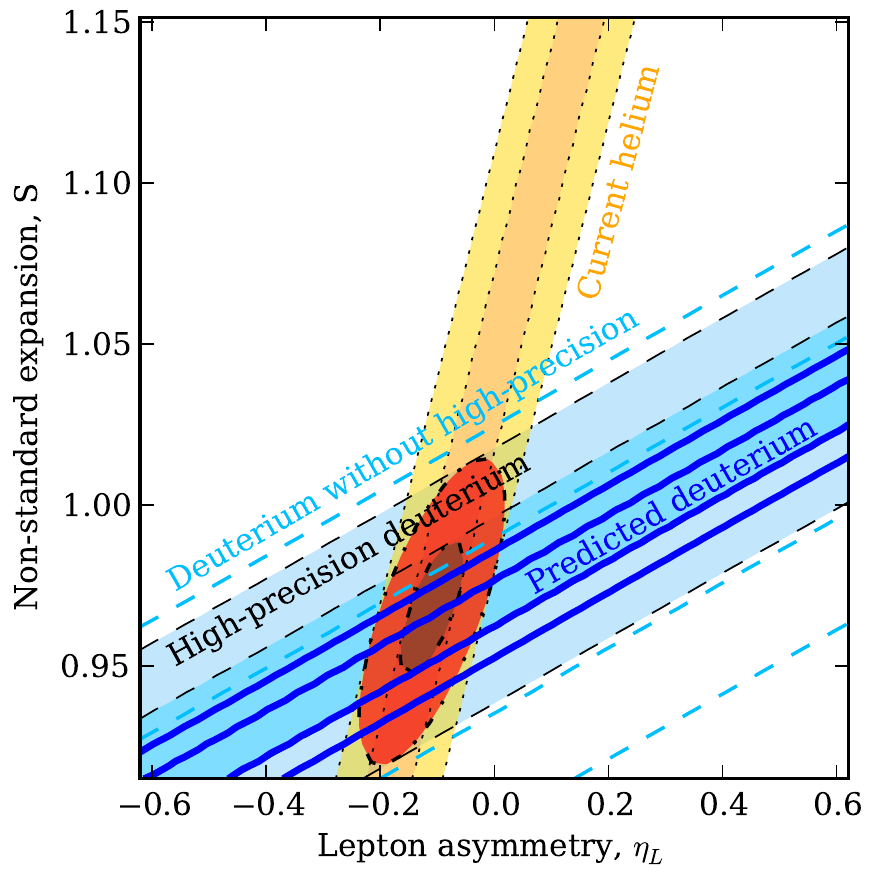}
\caption{Illustration of current and potential future constraints from nucleosynthesis on non-standard physics: $S$ is non-standard expansion as defined in \eqnref{S} and $\eta_L$ is the lepton asymmetry. The helium measurement is from \citet{Izotov:2013} and the dashed blue lines indicated the constraints from the weighted average of all measurements in \tabref{measurements} without the \citet{Cooke:2014} sample, while the shaded area is the constraint from \citet{Cooke:2014} alone, and the solid blue lines indicate the possible precision where the uncertainty is dominated by current limits on the nuclear reaction rates.}
\label{fig:future}
\end{figure}

\citet{Cooke:2014} presented a new measurement and four re-measurements of the deuterium abundance with very small uncertainties ($\approx 4$ per cent) using limiting selection criteria for the absorption systems with $\log (N(\mathrm{H}_\mathrm{I}) \geq 19$, minimum separation of $500\kms$ to other absorption systems, at least one unblended optically thin \Di{} transition available, several unblended metal lines with a range of oscillator strength available, high resolution data with resolution $R\geq 30000$ and $S/N>10$ per pixel available) and a ``blind'' analysis.\footnote{Blind in the sense that no model parameters are viewed until the fitting process has been finalised.} This demonstrates the potential of using deuterium as a high-precision probe of cosmological parameters, but the challenge of deciphering the structure of individual systems and the continuum may lead to underestimation of the uncertainties. E.g. the $z=3.049$ absorption system in \citet[][re-analysed in \cite{Cooke:2014}]{Pettini:2012} is redshifted so that it falls exactly on top of the  Lyman $\alpha$--N{\sc v} blend emission from the quasar. To avoid problems like this we need to perform a large number of deuterium measurements and look for a plateau for the primordial value, rather than rely on few values with a possible intrinsic scatter. While the strict criteria of \citet{Cooke:2014} do provide high precision measurements, in this paper we have shown that it is possible obtain robust measurements even if we relax the selection criteria and thereby increase the number of usable absorption systems significantly. \figref{future} shows current and predicted future constraints on non-standard nucleosynthesis from primordial abundances.

\subsection{The baryon fraction}
\Di/\Hi{} can be converted to \omegab{} using fitting formulae to Big Bang Nucleosynthesis (BBN) simulations \citep{Simha:2008b,Steigman:2007,Steigman:2012} 
\begin{equation}
10^5\mathrm{(D/H)} = K\left[ \frac{6}{\eta_{10}-6(S-1)}\right]^{1.6} \, ,
\end{equation}
where $\eta_{10} = 10^{10}n_\mathrm{b}/n_\gamma = 273.9 \Omega_\mathrm{b}h^2$ is the ratio of the number density of baryons to photons after electron-positron annihilation, the constant $K=2.55\pm0.08$ depends on the nuclear reaction rates \citep{Steigman:2012} and the 3 per cent uncertainty reflects the measured uncertainties in the nuclear reaction rates, in particular the D(p,$\gamma$)$^3$He cross section \citep[e.g.][]{Nollett:2011}. $S$ allows for non-standard radiation e.g. such as additional degrees of freedom $N_\mathrm{eff}>3.04$ through the radiation density $\rho_R'$,
\begin{equation} \label{eqn:S}
S = \frac{H'}{H}=\left(\frac{\rho '_R}{\rho_R} \right)^{1/2} = \left(1+\frac{7\Delta N_\nu}{43} \right)^{1/2} \, .
\end{equation}
$\Delta N_\nu$ is the parametrisation of extra radiation in terms of extra neutrino species defined as $N_\nu = 3+\Delta N_\nu$ \citep[see e.g.][]{Riemer-Sorensen:2013r}. Re-arranging the terms we get
\begin{equation}
\Omega_\mathrm{b}h^2 = 0.02191 \left[\left(\frac{10^5 \mathrm{(D/H)}}{K} \right)^{-0.625} +S -1\right] \, .
\end{equation}
For the standard BBN scenario with $S=1$ the observed values of \omegab{} are given in \tabref{measurements}.

If \Di/\Hi{} can be measured to a few percent uncertainty, we will reach a limit where the limiting factor are the nuclear reaction rates. As illustrated in \figref{future} it will be possible to use \Di/\Hi{} as a probe of non-standard physics even if this theoretical limit is not improved, if we can beat down the statistical and systematical uncertainty on the observed \Di/\Hi ~measurements.

\section{Conclusions} \label{sec:Conclusions}
The presented analysis of high quality spectra from multiple telescopes for the $z=3.256$ absorption system in J193957-100241 leads to a number of conclusions:
\begin{itemize}

\item A \Di/\Hi\ value of $2.45\pm 0.28$ corresponding to \omegab\ = $2.25\pm0.16\times10^{-5}$ for standard Big Bang nucleosynthesis. The main result was derived with the gas temperature and turbulent broadening as free parameters and fitting for summed column densities. The result is consistent if assuming turbulent broadening only (see \tabref{results}).

\item Given the large discrepancy between the previous measurement and the one presented here\footnote{$3\sigma$ if naively interpreted as a statistical fluctuation even though it origins in a systematical model difference.}, we conclude that data quality has a major impact. New and higher quality spectra exist for several of the absorption systems where \Di/\Hi\ has previously been measured and they should be re-visited.

\item We cross check the results between the telescopes and exclude instrument/telescope systematics as a source of scatter in the existing measurements.

\item We can obtain high precision measurements from lower column density systems than what is required by \citet{Cooke:2014}. The spread in column density could be important to understand the scatter among the measurements. Ideally we need a large sample of measurements to identify a plateau of primordial values.

\item High-precision measurements of the primordial abundances are a very important cross-check on standard model BBN, and allow us to probe models with non-standard physics. Even though the baryon density can be inferred from high-precision measurements of the CMB, the CMB probes the conditions at the time of recombination (approximately 300,000 years after Big Bang), whereas the primordial abundances probe the epoch of BBN (few minutes) directly. Beyond-standard model physics (e.g.~dark matter, extra relativistic particle species etc.) allows for different conditions between the two epochs \citet{Steigman:2013}, and consequently it is very important to observe both epochs independently.

\end{itemize}

\section*{Acknowledgments}
This research is based on observations
collected at the European Organisation for Astronomical Research in the
Southern Hemisphere, Chile, proposal ID 077.A-0166 obtained by PIs
Carswell, Kim, Haehnelt and Zaroubi.  It is also based on observations
collected with the Keck Observatory Archive (KOA), which is operated by
the W. M. Keck Observatory and the NASA Exoplanet Science Institute
(NExScI), under contract with the National Aeronautics and Space
Administration. The Keck data was obtained by PIs Songaila, Cowie,
Crighton and Tytler.  We are most grateful to an anonymous referee for a
timely and considerate report.  MTM thanks the Australian Research
Council for \textsl{Discovery Project} grant DP130100568 which supported
this work. Parts of this research were conducted by the Australian
Research Council Centre of Excellence for All-sky Astrophysics
(CAASTRO), through project number CE110001020.

\bibliographystyle{mn2e}
\bibliography{bibliography}

\appendix
\section{Model} \label{app:Model}

\begin{table*}
\centering
\begin{minipage}{140mm}
\label{tab:model}
\caption{The model. The Keck observations from \tabref{data} are stacked into setup 2, setup 5 and setup 10, and all the VLT observations are stacked into one spectrum. The b-parameters are thermally tied within the listed groups except when marked otherwise.}\label{tab:allmodel}
\begin{tabular}{l l l r r r }
\hline
Component		& Redshift 				& Specie		& $log(N)$			& $b_\mathrm{turb}$ or $b_\mathrm{tot}$ $[\kms]$ 	& $T$ [$10^{4} ~\mathrm{K}$]\\ \hline
A				& $3.25576\pm0.000008$		& \Siiv		& $11.349\pm0.002$		& $2.83\pm0.80$\footnotemark[1]	& ---	\\
				&						& \Civ		& $11.970\pm0.008$		& $4.32\pm0.00$\footnotemark[1]	& ---	 \\
				&						& \Cii		& $11.911\pm0.032$		& $4.93\pm8.03$				& $2.02\pm0.78$ \\
				&						& \Oi			& $12.794\pm0.000$\footnotemark[3] & $4.93\pm8.03$		& $2.02\pm0.78$ \\
				&						& \Hi			& $15.359\pm 0.651$	& $4.93\pm8.03$				& $2.02\pm0.78$ \\
				&						& \Di			& $10.750\pm0.000$\footnotemark[3] & $4.93\pm8.03$		& $2.02\pm0.78$ \\

B 				& $3.25593\pm0.000003$		& \Siiv		& $11.827\pm0.082$		& $5.51\pm1.33$				& --- \\
				&						& \Cii		& $13.364\pm0.025$		& $2.03\pm0.62$				& $1.50\pm0.08$ \\
				&						& \Siii		& $12.505\pm0.017$		& $2.03\pm0.62$				& $1.50\pm0.08$ \\
				&						& \Alii		& $11.482\pm0.014$		& $2.03\pm0.62$				& $1.50\pm0.08$ \\
				&						& \Oi			& $12.604\pm0.000$\footnotemark[3] & $2.03\pm0.62$		& $1.50\pm0.08$ \\
				&						& \Hi			& $17.898\pm0.045$		& $2.03\pm0.62$				& $1.50\pm0.08$ \\
				&						& \Di			& $13.289\pm0.000$\footnotemark[3] & $2.03\pm0.62$		& $1.50\pm0.08$ \\

C				& $3.25603\pm0.000002$ 	& \Siiv		& $12.734\pm0.014$		& $5.93\pm0.24$\footnotemark[1]	& --- \\
				&						& \Civ		& $12.755\pm0.016$		& $9.05\pm0.00$\footnotemark[1]	& --- \\
				&						& \Cii		& $13.107\pm0.051$		& $3.00\pm1.05$				& $1.03\pm0.45$ \\
				&						& \Siii		& $12.056\pm0.053$		& $3.00\pm1.05$				& $1.03\pm0.45$ \\
				&						& \Alii		& $11.269\pm0.046$		& $3.00\pm1.05$				& $1.03\pm0.45$ \\
				&						& \Oi			& $11.841\pm0.000$\footnotemark[3] & $3.00\pm1.05$		& $1.03\pm0.45$ \\
				&						& \Hi			& $17.135\pm0.134$		& $3.00\pm1.05$				& $1.03\pm0.45$ \\
				&						& \Di			& $12.526\pm0.000$\footnotemark[3] & $3.00\pm1.05$		& $1.03\pm0.45$ \\

D				& $3.25626\pm0.000003$		& \Siiv		& $12.284\pm0.016$		& $7.45\pm0.37$\footnotemark[1]	& --- \\
				&						& \Civ		& $12.264\pm0.041$		& $11.40\pm0.00$\footnotemark[1]	& --- \\
				&						& \Siii		& $11.941\pm0.100$		& $12.29\pm2.30$			 	& $2.47\pm0.36$ \\
				&						& \Alii		& $10.850\pm0.109$		& $12.29\pm2.30$			 	& $2.47\pm0.36$ \\
				&						& \Oi			& $11.702\pm0.000$\footnotemark[3] & $12.29\pm2.30$		& $2.47\pm0.36$ \\
				&						& \Hi			& $16.997\pm0.033$		& $12.29\pm2.30$			 	& $2.47\pm0.36$ \\
				&						& \Di			& $12.388\pm0.000$\footnotemark[3] & $12.29\pm2.30$		& $2.47\pm0.36$ \\

E				& $3.25659\pm0.000002$		& \Siiv	 	& $12.084\pm0.012$		& $7.36\pm0.31$\footnotemark[2]	& ---\\
				&						& \Civ		& $12.110\pm0.037$		& $7.36\pm0.31$\footnotemark[2]	& --- \\
				&						& \Oi			& $11.999\pm0.000$\footnotemark[3] & $7.36\pm0.31$\footnotemark[2]	& --- \\
				&						& \Hi			& $17.294\pm0.192$		& $7.36\pm0.31$\footnotemark[2]	& --- \\
				&						& \Di			& $12.685\pm0.000$\footnotemark[3] & $7.36\pm0.31$\footnotemark[2]	& --- \\
\hline

Summed 			&						& Specie		& $\sum\log(N)$ \\ \hline
				&						& \Siiv		& $12.976\pm0.002$ \\
				&						& \Civ		& $12.989\pm0.009$ \\				
				&						& \Siii		& $12.717\pm0.017$ \\
				&						& \Cii		& $13.536\pm0.037$ \\
				&						& \Alii		& $11.748\pm0.015$ \\
				&						& \Oi			& $12.794\pm0.171$ \\
				&						& \Hi			& $18.088\pm0.035$ \\
				&						& \Di			& $13.479\pm0.043$ \\
\end{tabular}
\vspace{1cm}

\footnotemark[1]{Thermally tied between \Civ{} and \Siiv{}}\\
\footnotemark[2]{Turbulently tied between \Civ{}, \Siiv{}, \Hi{} and \Di{}}\\
\footnotemark[3]{Ratio tied to \Hi}
\end{minipage}
\end{table*}

\begin{table*}
\centering
\begin{minipage}{140mm}
\contcaption{The model}
\begin{tabular}{l l l l r r r}

\hline
Blends 			& Redshift 	& Specie		& $log(N)$	& b $[\kms]$ 	& $T$ [$10^{4} ~\mathrm{K}$]\\ \hline
\Cii{} 1334	\AA		& 2.62638		& \Hi    		& 13.000   	& 25.07 		& --- & \\
				& 2.62748		& \Hi    		& 14.053   	& 40.42 		& --- & \\
 				& 2.62741		& \Hi    		& 13.189   	& 11.71 		& --- & \\
				& 2.62770		& \Hi    		& 13.029   	& 11.32 		& --- & \\
				& 2.62826		& \Hi    		& 12.524   	& 13.40 		& --- & \\
				& 2.62884		& \Hi    		& 13.804   	& 28.45 		& --- & \\
 				& 2.62926		& \Hi    		& 13.368   	& 19.32 		& --- & \\
				& 2.63050		& \Hi    		& 14.376   	& 24.45 		& --- & \\

\Cii{} 1036 \AA		& 3.67060		& \Hi			& 12.884 		& 10.31	& ---\\
				& 3.67084  	& \Hi			& 13.442		& 15.46	& ---\\
				& 3.67293  	& \Hi			& 14.243		& 18.73	& ---\\
				& 3.67235		& \Hi			& 13.064		& 22.47	& ---\\  
				& 3.67357   	& \Hi			& 12.509		& 18.84	& ---\\
				& 3.67447    	& \Hi			& 14.058		& 35.56	& ---\\
				& 3.67437    	& \Hi			& 13.375		& 20.72	& --- \\ 
				& 3.67511    	& \Hi			& 13.672		& 18.31	& ---\\  
					
\Siii{} 1526 \AA		& 3.41150    	& \Hi			& 13.062		& 21.61	& --- \\
				& 3.41217    	& \Hi			& 13.586		& 21.48	& --- \\
				& 3.41241    	& \Hi			& 13.290		& 25.04 	& --- \\
				& 3.41311    	& \Hi			& 13.499		& 27.81	& --- \\
												
\Hi{} 1216 \AA		& 2.59729		& \Hi			& 12.176		&  5.67 & --- \\
				& 2.59712		& \Hi			& 12.751	  	&  8.83  & --- \\
				& 0.56019		& \Mgii		& 11.837		& 13.73 	& --- \\
				& 0.56030		& \Mgii		& 13.620		& 22.00 	& --- \\
				& 2.59902		& \Hi			& 12.692		& 25.43 	& --- \\
				& 2.58590 	& \Hi 		& 12.500 		& 20.27 	& --- \\
				& 2.58638		& \Hi 		& 12.257		& 10.18 	& --- \\
				& 2.58736		& \Hi 		& 13.186		& 58.58 	& --- \\
				& 2.58734		& \Hi 		& 13.388		& 24.43  & --- \\
				& 2.58777		& \Hi 		& 12.338		&  6.03 & --- \\					
				& 2.58854		& \Hi			& 13.210		& 35.19 	& --- \\
				& 2.58878		& \Hi 		& 13.649		& 18.10 	& --- \\
				& 2.58944		& \Hi 		& 13.695		& 18.54 	& --- \\
				& 2.59243		& \Hi			& 12.710		& 20.65 	& --- \\
				& 2.59305		& \Hi 		& 14.076		& 30.07 	& --- \\
				& 2.59355		& \Hi 		& 12.860		& 13.58 	& --- \\
				& 2.59392		& \Hi 		& 13.155		& 21.44 	& --- \\
					
\Hi{} 1026 \AA	& 3.25253		& \Hi 		& 12.196		& 29.99	& --- \\
				& 3.25405		& \Hi 		& 13.446		& 47.00	& --- \\
				& 3.25390		& \Hi 		& 13.287		& 21.34	& --- \\
				& 3.25754		& \Hi		& 12.865		& 23.12	& --- \\
				& 3.25824		& \Hi 		& 12.938		& 19.37	& --- \\
				& 3.25889		& \Hi 		& 12.613		& 34.09	& --- \\
				& 3.25935		& \Hi 		& 12.154		& 17.39	& --- \\
				& 3.26008		& \Hi 		& 12.824		& 14.57	& --- \\
				& 3.26030		& \Hi 		& 13.255		& 28.76	& --- \\
				& 3.26193		& \Hi 		& 13.079		& 25.18	& --- \\
				& 3.26238		& \Hi 		& 12.459		& 22.93	& --- \\
				& 3.26368		& \Hi 		& 12.404		&  5.83	& --- \\

\hline

Velocity shift		&			& Data		& Shift [$\kms$] \\ \hline
				&			& setup 2		& --- \\
				&			& setup 5		& $0.344\pm0.068$ \\
				&			& setup 10 \Siiv	& $-0.008\pm0.110$\\
				&			& setup 10 	& $-0.785\pm0.068$ \\
				&			& VLT		& $0.126\pm0.062$ \\

\end{tabular}
\\

\end{minipage}
\end{table*}

\bsp

\label{lastpage}

\end{document}